\documentclass[12pt]{article}
\usepackage{axodraw}
\usepackage{epsfig}   
\usepackage{amssymb}
\usepackage{latexsym}
\usepackage{amsmath}

\usepackage{graphics,subfigure} 

\usepackage{color}   

\newcommand{\bea}{\begin{eqnarray}}
\newcommand{\eea}{\end{eqnarray}}
\newcommand{\beq} {\begin{equation}}
\newcommand{\eeq} {\end{equation}}

\begin{document}
\pagestyle{empty}
\begin{flushright}
\today
\end{flushright}
\begin{center}
{\large\sc {\bf A short note on Magnetized Black-hole in Non-linear Electrodynamics }}\\
\end{center}
\vspace{1.0truecm}
\begin{center}
{\large  ~H.~A.~Redekar,~R.~B.~Kumbhar,~S.~P.~Das \footnote{Email: spd.phy@unishivaji.ac.in} and K.~Y.~Rajpure }\\
\vspace*{5mm}
{}{\it                                                                      
Department of Physics, Shivaji University, \\[0.15cm]
Kolhapur-416004, Maharashtra, India.} \\[0.07cm]
\end{center}

\begin{abstract}
We have analyzed the thermodynamic properties of magnetized black-hole in the 
background of non-linear electrodynamics with two parameters $\beta$ and $\gamma$. We have studied the Bekenstein-Hawking entropy, Hawking temperature,
specific heats in two-dimesional surface plots as a function of event horizon ($r_{+}$) and $\gamma$. We showed the variation profiles of the above thermodynamic parameters for $\gamma$ [$ 0 \rightarrow 1$]. We identified regions of parameters for the possible phase-transitions and the stability of the black-holes.
\end{abstract}

\newpage
\setcounter{page}{1}
\pagestyle{plain}

\section{Introduction}
\label{sec:intro}

The solutions of Black-hole (BH) are natural outcome of the General theory of Relativity (GTR). Generally 
the BH characterized by their radius, masses, charges, angular momentum etc. Depending upon the parameter 
choices BH could have very different types, e.g., the  Schwarzschild, Kerr, Kerr-Newman, Reissner 
Nordstr\"{o}m\cite{Jacobson:2012ei}, \cite{Curiel:2018cbt}, \cite{Miller:2014aaa}.

Non-linear electrodynamics (NLED) has been found useful in astrophysics. 
With the inclusion of GTR we called the combined theory as NLED-GTR. It turns out that initial singularities in early 
universe are absent in NLED-GTR model. This theory has some remarkable features, e.g., 
consistent theory of inflationary model of the Universe 
\cite{Garcia,Camara,Elizalde,Novello,Novello1,Vollick,Kruglov3}. This NLED-GTR model 
has few characteristic features, e.g., absence of initial singularities, putting  an 
upper limit on the Electric field at the origin of point-like particles, the finite 
self-energy of the charged particles \cite{Novello1}, \cite{Born} \cite{Kruglov1}. 
Due to loop corrections the non-linear terms arises  in Quantum Electrodynamics(QED) 
\cite{Heisenberg, Schwinger, Adler}. In the model under considerations correspondence 
principle holds good. Therefore, NLED turns into Maxwell’s equations in the weak field limit.

The physics of BH (both electrically and magnetically charged) in presence of NLED has 
been studied since some time \cite{Kruglov1}, \cite{Bardeen}, \cite{Dymnikova}, \cite{Mazharimousavi:2021uki}, \cite{Shabad2}. 
We recently studied the thermodynamic properties of slowly rotating magnetized black-hole \cite{Managave:2023rhn} 
in NLED-GTR scenario. In this  analysis we consider no rotation, i.e., purely static Schwarzschild type of 
BH. In our present analysis we follow \cite{Kruglov1} but extended the idea by choosing the continuous values of $\gamma$ parameter in the NLED-GTR model.

This paper is organised as follows. We briefly outline the NLED-GTR model in section 2.  
We used the model of NLED with two parameters $\beta$ and $\gamma$ and  deduced the magnetized mass density that affects the relevant thermodynamic parameters. 
Then we go through asymptotic of metric and mass function at $r \to 0$ and  $r \to \infty$ for continuous variations of $\gamma$-parameter in the ranges between 0 to 1. The black hole thermodynamics, i.e.,  the entropy discussed along with 
Hawking temperature and heat capacity of black holes. In section 5 we discussed summary and the conclusion.

We consider natural units, i.e., $c=\hbar=1$, $\varepsilon_0=\mu_0=1$, and 
metric signature $\eta=\mbox{diag}(-1,1,1,1)$

\section{ Non-linear Electrodynamics (NLED)-model}

The Lagrangian density of NLED model is as (following \cite{Kruglov1}): 
\begin{equation}
{\cal L} = -\frac{{\cal F}}{1+(\beta{\cal F})^\gamma},
 \label{lag1}
\end{equation}
where the parameter $\beta$ has the dimensions of 4D-volume (manifests the strength of the coupling), and $\gamma$ (manifests the order of interaction) is the dimensionless parameter \cite{Kruglov1}. The ${\cal F}=(1/4)F_{\mu\nu}F^{\mu\nu}=(\textbf{B}^2-\textbf{E}^2)/2$, where 
$F_{\mu\nu}=\partial_\mu A_\nu-\partial_\nu A_\mu$ is the field strength tensor, with $A_{\mu}= (\phi, \vec A)$ the $\phi$ and $\vec A$ are the scalar and vector potential respectively. 

By using the Euler-Lagrange (EL) equation one can obtain the field equations \begin{equation} 
\partial_\mu\left({\cal L}_{\cal F}F^{\mu\nu} \right)=0,
\label{lag2}
\end{equation} 
where ${\cal L}_{\cal F}=\partial {\cal L}/\partial{\cal F}$.

The EL equation on Lagrangian density leads to \begin{equation}\label{lag3}
  {\cal L}_{\cal F}=\frac{(\gamma-1)(\beta{\cal F})^\gamma-1}{(1+(\beta{\cal F})^\gamma)^2}.
\end{equation}

To hold the causality principle, the group velocity over the background is less than photon 
speed to be $ {\cal L}_{\cal F}\leq 0$ \cite{Shabad2}. This confirms the absence of tachyon. 
In our analysis we imply that $0\leq \gamma \leq 1$. We considered only magnetized black holes 
(\textbf{E}=0, ${\cal F}=\textbf{B}^2/2$). 

The symmetrical energy-momentum tensor (is obtained from Eqn. \ref{lag1} ) is: 

\begin{equation}
T_{\mu\nu}=\frac{(\gamma-1)(\beta{\cal F})^\gamma-1}{[1+(\beta{\cal F})^\gamma]^2} F_\mu^{~\alpha}F_{\nu\alpha}
-g_{\mu\nu}{\cal L}.
\label{trace1}
\end{equation}

The trace of the energy-momentum tensor Eqn.\ref{trace1} leads to 

\begin{equation}\label{10}
  {\cal T}\equiv T^{\mu}_\mu=\frac{4\gamma {\cal F}(\beta{\cal F})^\gamma}{[1+(\beta{\cal F})^\gamma]^2}.
\end{equation}

The trace of Eqn.5 is non-zero due to the presence of dimensional parameter $\beta=1$ and scale invariance is broken.

The models exhibits Maxwell's Electrodynamics at the weak field limit, $\beta {\cal F} \ll 1$, ${\cal L}\rightarrow-{\cal F}$, i.e., holds the correspondence principle.

\section{Magnetized black holes}

We will study magnetically charged black hole ($\textbf{E}=0$). The action of NLED-GTR is

\begin{equation}
I=\int d^4x\sqrt{-g}\left(\frac{1}{2\kappa^2}R+ {\cal L}\right),
\label{action}
\end{equation}

where $\kappa^2=8\pi G\equiv M_{Pl}^{-2}$, $G$ is Newton's constant, $M_{Pl}$ is the reduced Planck mass, 
and $R$ is the Ricci scalar. By varying action Eqn.\ref{action} with respect to the metric and 
electric potential one can obtain the Einstein and electromagnetic field equations

\begin{equation}
R_{\mu\nu}-\frac{1}{2}g_{\mu\nu}R=-\kappa^2T_{\mu\nu},
\label{gtrpartial}
\end{equation}

\begin{equation}
\partial_\mu\left[\sqrt{-g}{\cal L}_{\cal F}F^{\mu\nu}\right]=0.
\label{empartial}
\end{equation}

The line element having the spherical symmetry is 

\begin{equation}
ds^2=-f(r)dt^2+\frac{1}{f(r)}dr^2+r^2(d\vartheta^2+\sin^2\vartheta d\phi^2).
\label{lineelement}
\end{equation}

The metric function is defined by the relation \cite{Bronnikov}
\begin{equation}
f(r)=1-\frac{2GM(r)}{r},
\label{metric}
\end{equation}

and the mass function is given by
\begin{equation}
M(r)=\int_0^r\rho_M(r)r^2dr,
\label{massfn}
\end{equation}
where $\rho_M$ is the magnetic energy density.

The magnetic energy density (\textbf{E}=0) is found from Eq.\ref{trace1}
\begin{equation}\label{rho}
  \rho_M=T_0^{~0}=\frac{{\cal F}}{1+(\beta{\cal F})^\gamma}.
\end{equation}

For the magnetized black hole \cite{Bronnikov} the field-strength with magnetic charge($q$) and at distance $r$ is ${\cal F}=q^2/(2r^4)$. For simplicity we set the value of $q=1$ and $\beta=1$ (which turns out 
to be some kind of scale factor) in our analysis.

\section{Thermodynamic Parameters} In this section we are summarizing few important thermodynamic parameters which are relevant to our analysis.\\

$\bullet$ The Bekenstein-Hawking entropy (area law) is 
\begin{equation}
	S_{BH} = \frac{\pi r_{+}^2 }{G} 
\end{equation}
where $r_{+}$ is the event horizon and G is the Newton's Gravitational constant. We set G=1 in our analysis.

$\bullet$ The Hawking temperature of the black hole horizon is
\begin{equation}
T_H=\frac{\kappa_S}{2\pi}=\frac{f'(r_+)}{4\pi}, 
\label{thexp}
\end{equation}
where $\kappa_S$ is the surface gravity.

$\bullet$ The specific heat capacity
\begin{equation}
C_q=T_H\left(\frac{\partial S}{\partial T_H}\right)_q=\frac{T_H\partial S/\partial r_+}{\partial T_H/\partial r_+}=\frac{2\pi r_+T_H}{G\partial T_H/\partial r_+}.
\label{specificheat}
\end{equation}
We will see that (following Eq.\ref{specificheat}) the heat capacity diverges if the Hawking temperature has extremum.

\section{Numerical analysis}

\begin{figure}[ht!]
\begin{center}
\raisebox{0.0cm}{\hbox{\includegraphics[angle=0,scale=0.35]{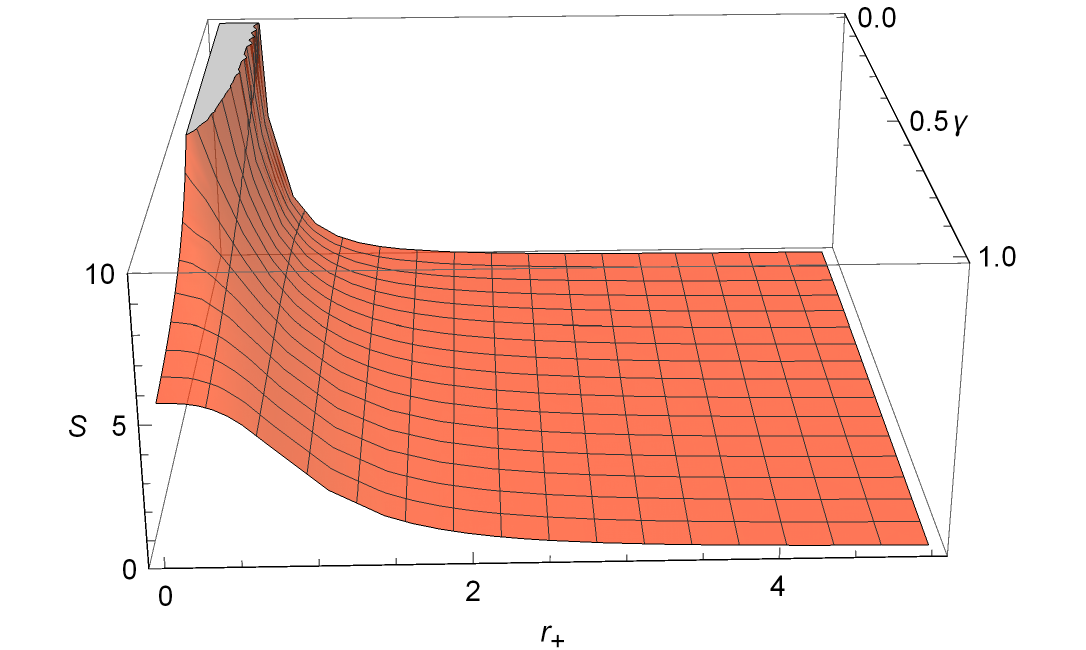}}}
\raisebox{0.0cm}{\hbox{\includegraphics[angle=0,scale=0.35]{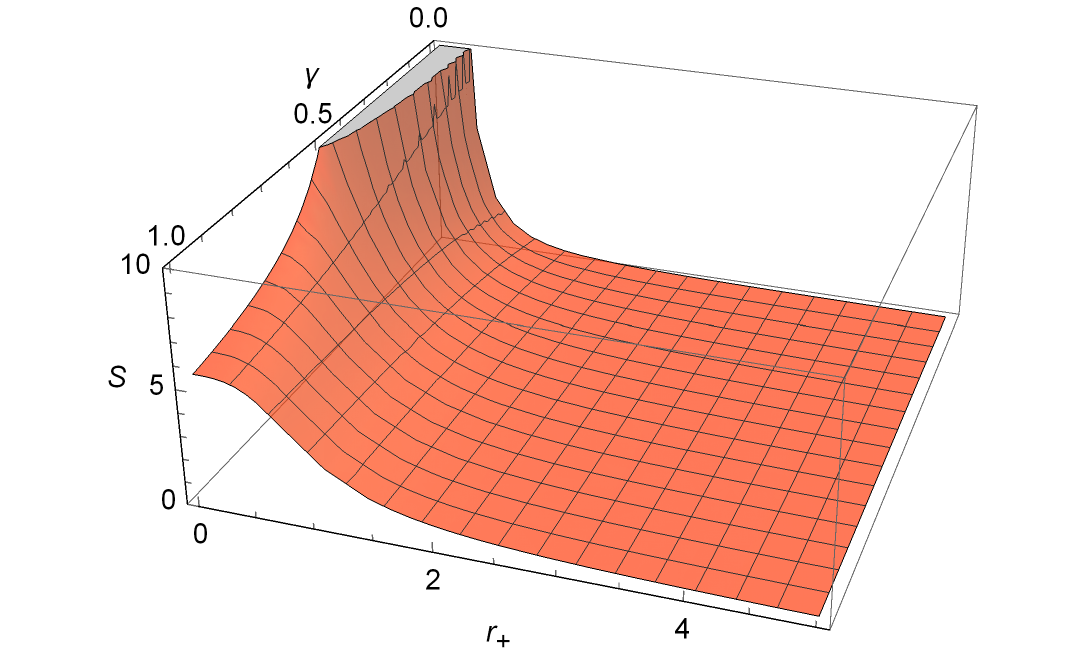}}}
	\caption{Entropy as a function of radial co-ordinates ($r_{+}$) and $\gamma$ in two different orientations}
\label{figrgs}
\end{center}
\end{figure}

In our numerical analysis we set $\beta=1$, $q=1$ and $\gamma$ as a free parameter and varied in the ranges between $0 \to 1$. In Fig.\ref{figrgs} we showed the Bekenstein-Hawking entropy as a function of radial co-ordinates ($r_{+}$) and $\gamma$.  

\begin{figure}[ht!]
\begin{center}
\raisebox{0.0cm}{\hbox{\includegraphics[angle=0,scale=0.38]{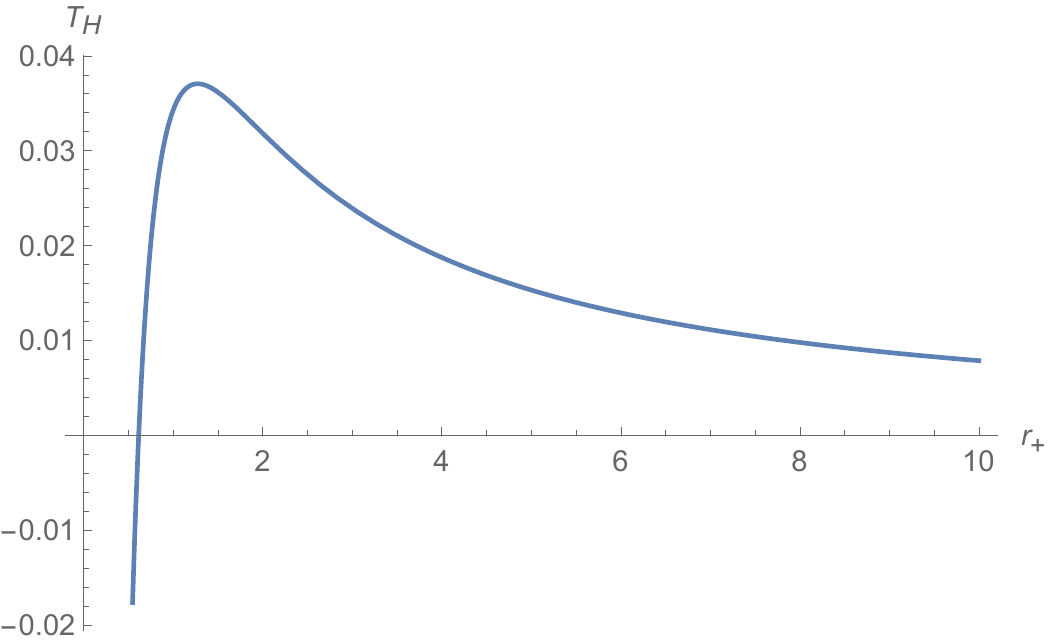}}}
\raisebox{0.0cm}{\hbox{\includegraphics[angle=0,scale=0.38]{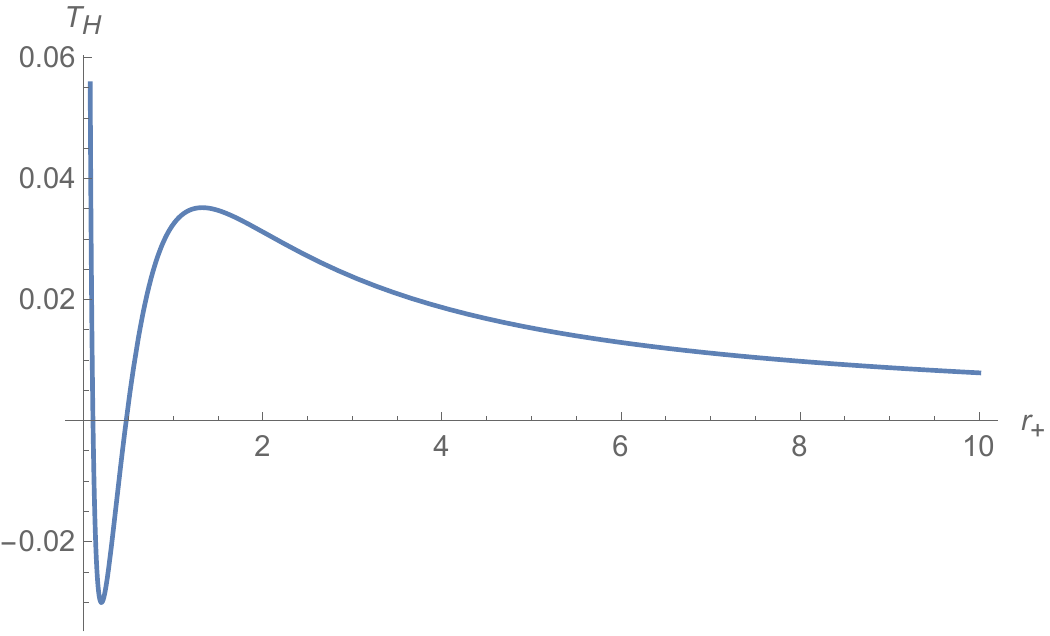}}}
\raisebox{0.0cm}{\hbox{\includegraphics[angle=0,scale=0.35]{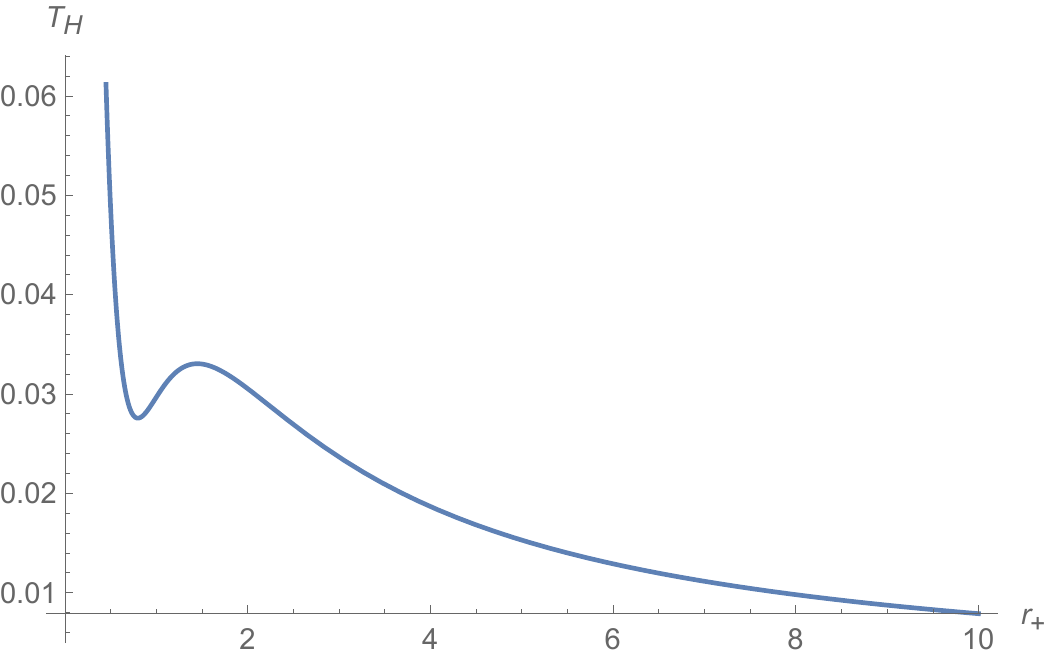}}}
\raisebox{0.0cm}{\hbox{\includegraphics[angle=0,scale=0.38]{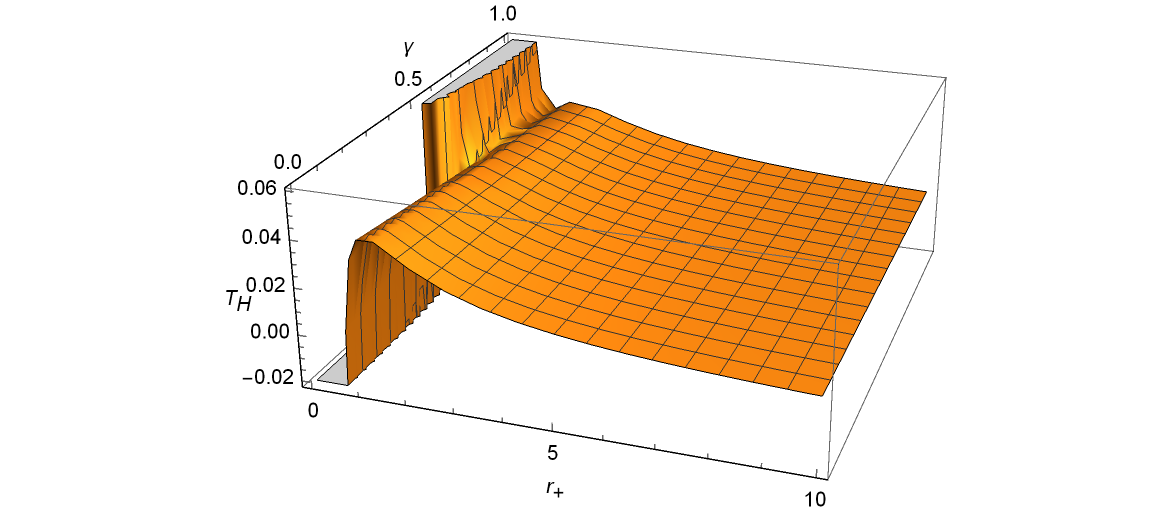}}}
\caption{The Hawking temperature as a function of $r_{+}$ for $\gamma$= $0.40, 0.54, 0.75$ and as a surface plots in $r_{+} - \gamma$ ( in bottom-right panel)}
\label{figth}
\end{center}
\end{figure}

The Hawking temperature is plotted in Fig.\ref{figth}. We know that as long as the Hawking temperature is positive the BH is stable. So from the Fig.\ref{figth} we see that the BH is stable for $r_{+}=0.51$ for $\gamma=0.40$ (top left panel). If we increase the value of $\gamma=0.54$ or bit larger then we found that in small interval of $r_{+}=[0.5-0.01]$ the BH is unstable. 

If we further increase the value of $\gamma=0.75$ (bottom-left panel) the unstability never arises. 
So large values of $\gamma$ is better for stability of the Black-hole. 

In the bottom-right panel of Fig.\ref{figth} we showed in 2d-surface plots of Hawking temperature in $r_{+}$-$\gamma$ plane.

We know that the first-order phase transition occurs when the Hawking temperature changes the sign. From Fig.\ref{figth} we see that for $\gamma=0.40$ amd 
$\gamma=0.54$ the first order phase transition take places. However once the value of 
$\gamma \geq 0.57$ the Hawking temperature never encounters negative values and hence first-order phase transitions cannot take places.

\begin{figure}[ht!]
\begin{center}
\raisebox{0.0cm}{\hbox{\includegraphics[angle=0,scale=0.375]{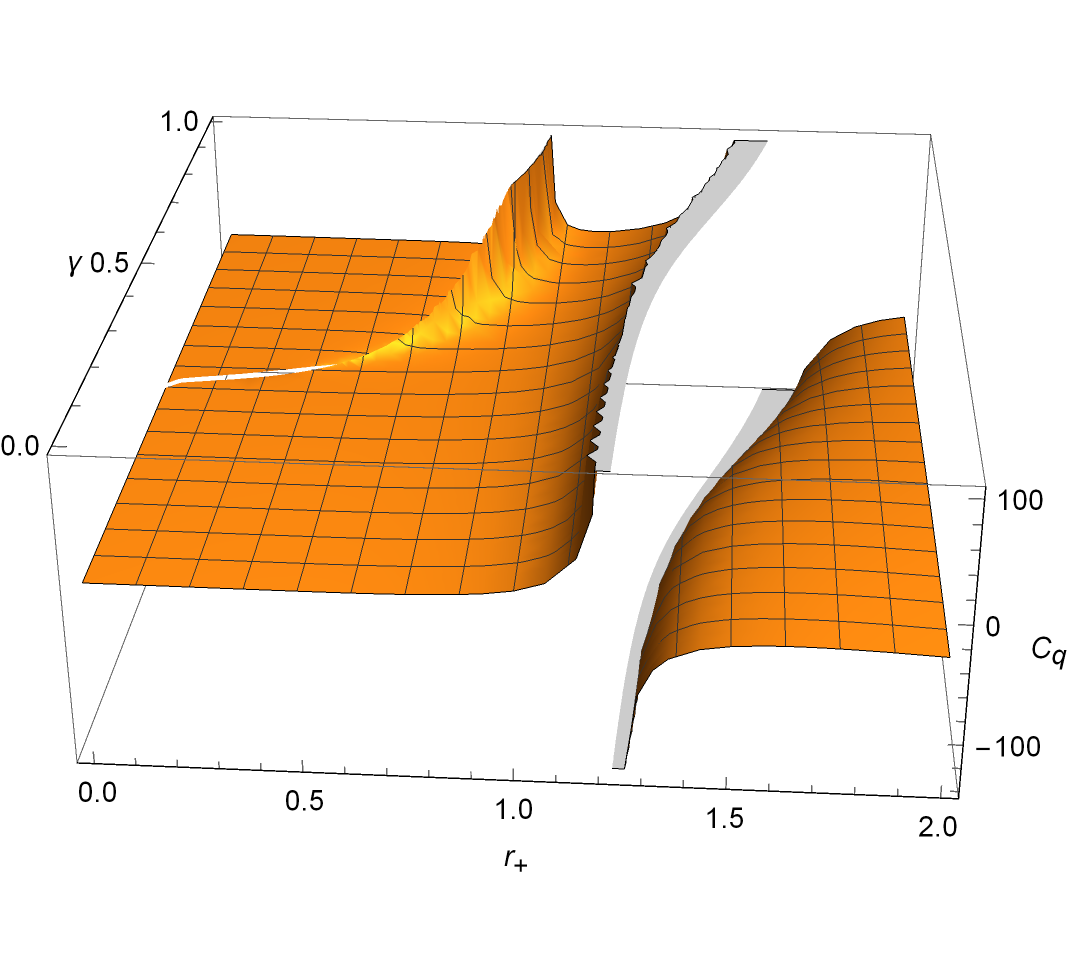}}}
\raisebox{0.0cm}{\hbox{\includegraphics[angle=0,scale=0.375]{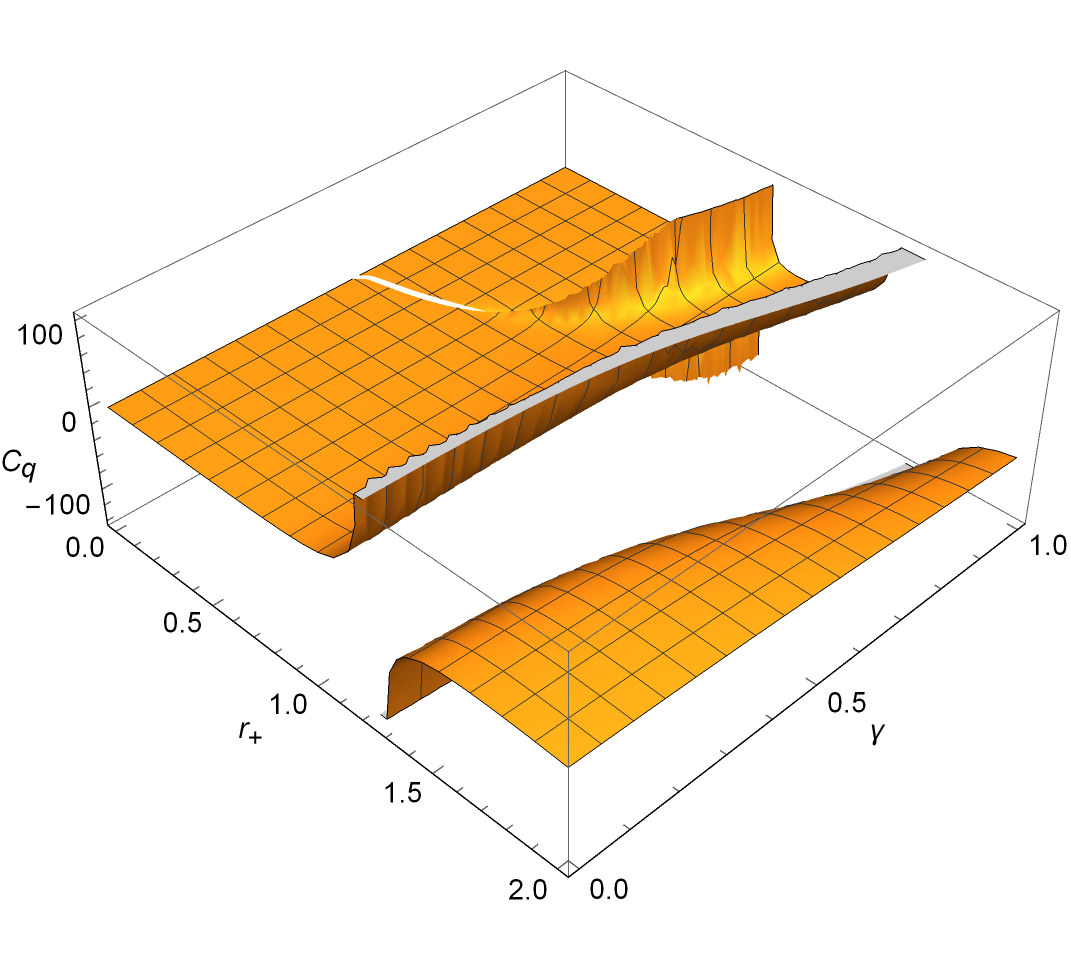}}}
	\caption{ The specific heat ($C_q$) as a function of $r_{+}$ and $\gamma$ in two different orientation.}
\label{figcq}
\end{center}
\end{figure}

We know that heat capacity diverges if the Hawking temperature has extremum. From the bottom-right panel of Fig.~\ref{figth} we see that heat capacity diverges around $r=1.4$ for almost all values of $\gamma$. 

We can also say from the bottom-right panel of Fig.\ref{figth} 
that around $r=1.4$ the second-order phase transition take 
places for all values for $\gamma$.

In Fig. \ref{figcq} we plotted the specific heat ($C_q$) as a function of $r_{+}$ and $\gamma$ in two 
differnt orientations. The specific heat is discontinuous in some wide-ranges around $r_{+}=1.1-1.6$ 
depending upon the values of $\gamma$. Large values of $\gamma$ leads large values of 
$r_{+}$. The BH undergoes second order 
phase-transition around this ranges. As seen from Fig.\ref{figcq} that 
in some wide ranges of $r_{+}=1.2$ to $ 1.5$ the heat capacity is negative (generically for larger values of $\gamma$ leads larger values of $r_{+}$) and in that parameter spaces the BH is unstable. 

\section{Summary and Conclusion}

We have analyzed the thermodynamic properties of magnetized Black-hole in non-linear electrodynamics model. The model is characterized by two parameters $\beta$ and $\gamma$. In our numerical analysis 
we set $\beta=1$ throughout. We varied $\gamma$-parameter in the ranges between $0 \to 1$ and analyze the Hawking temperature, black-hole entropy, specific-heat capacity and stability. We have demarcated regions using $r_{+}-\gamma$ surface plots where the Black-holes undergo second-order phase transitions and regions of instability.
\vspace{6pt} 


%

\begin{thebibliography}{999}
\bibitem{Jacobson:2012ei}
T.~Jacobson,
Lect. Notes Phys. \textbf{870}, 1-29 (2013)
doi:10.1007/978-3-319-00266-8\_1
[arXiv:1212.6821 [gr-qc]].

\bibitem{Curiel:2018cbt}
E.~Curiel,
Nature Astron. \textbf{3}, no.1, 27-34 (2019)
doi:10.1038/s41550-018-0602-1
[arXiv:1808.01507 [physics.hist-ph]].

\bibitem{Miller:2014aaa} M.~C.~Miller and J.~M.~Miller,
Phys. Rept. \textbf{548}, 1-34 (2014)
doi:10.1016/j.physrep.2014.09.003
[arXiv:1408.4145 [astro-ph.HE]].

\bibitem{Garcia} R. Garc\'{i}a-Salcedo and N. Breton, Int. J. Mod. Phys. A \textbf{15}, 4341 (2000) [arXiv:gr-qc/0004017].
\bibitem{Camara} C. S. Camara, M. R. de Garcia Maia, J. C. Carvalho and J. A. S. Lima, Phys. Rev. D \textbf{69}, 123504 (2004) [arXiv:astro-ph/0402311].

\bibitem{Elizalde} E. Elizalde, J. E. Lidsey, S. Nojiri and S. D. Odintsov, Phys. Lett. B \textbf{574}, 1 (2003) [arXiv:hep-th/0307177].

\bibitem{Novello} M. Novello, S. E. Perez Bergliaffa and J. M. Salim, Phys. Rev. D \textbf{69}, 127301 (2004) [arXiv:astro-ph/0312093].

\bibitem{Novello1} M. Novello, E. Goulart, J. M. Salim and S. E. Perez Bergliaffa, Class. Quant. Grav. \textbf{24}, 3021 (2007) [arXiv:gr-qc/0610043].

\bibitem{Vollick} D. N. Vollick, Phys. Rev. D \textbf{78}, 063524 (2008) [arXiv:0807.0448].

\bibitem{Kruglov3} S. I. Kruglov, Phys. Rev. D \textbf{92}, 123523 (2015) [arXiv:1601.06309];
Int. J. Mod. Phys. A \textbf{31}, 1650058 (2016) [arXiv:1607.03923];
Int. J.Mod. Phys. D \textbf{25}, 1640002 (2016) [arXiv:1603.07326].

\bibitem{Born} M. Born and L. Infeld, Proc. Royal Soc. (London) A \textbf{144}, 425 (1934).


\bibitem{Kruglov1} S.~I.~Kruglov,
Annalen Phys. \textbf{529}, no.8, 1700073 (2017)
doi:10.1002/andp.201700073
[arXiv:1708.07006 [gr-qc]].

\bibitem{Heisenberg} W. Heisenberg and H. Euler, Z. Physik, \textbf{98}, 714 (1936) [arXiv:physics/0605038].

\bibitem{Schwinger} J. Schwinger, Phys. Rev. \textbf{82}, 664 (1951).

\bibitem{Adler} S. L. Adler, Ann. Phys. (N.Y.) \textbf{67}, 599 (1971).

\bibitem{Bardeen} J. M. Bardeen, in Proc. Int. Conf. GR5, Tbilisi, p. 174, 1968.

\bibitem{Dymnikova} I. Dymnikova, Gen. Rev. Grav. \textbf{24}, 235 (1992); E. Ay\'{o}n-Beato, A. Gar\'{c}ia, Phys. Rev. Lett.  \textbf{80}, 5056 (1998)
[arXiv:gr-qc/9911046]; K. A. Bronnikov, Phys. Rev. D \textbf{63}, 044005 (2001);N. Breton, Phys. Rev. D \textbf{67}, 124004 (2003) [arXiv:hep-th/0301254]; S. A. Hayward, Phys. Rev. Lett. \textbf{96}, 31103 (2006) [arXiv:gr-qc/0506126]; J. P. S. Lemos and V. T. Zanchin, Phys. Rev. D \textbf{83}, 124005 (2011)
[arXiv:1104.4790]; A. Flachi and J. P.S. Lemos, Phys. Rev. D \textbf{87}, 024034 (2013)
[arXiv:1211.6212]; S. H. Hendi, Ann. Phys. \textbf{333}, 282 (2013) [arXiv:1405.5359]; L. Balart and E. C. Vagenas, Phys. Rev. D \textbf{90}, 124045 (2014) [arXiv:1408.0306]; S. I. Kruglov, Phys. Rev. D \textbf{94}, 044026 (2016) [arXiv:1608.04275]; Europhys. Lett. \textbf{115}, 60006 (2016) [arXiv:1611.02963]; Ann. Phys. (Berlin) \textbf{528}, 588 (2016) [arXiv:1607.07726]; V. P. Frolov, Phys. Rev. D \textbf{94}, 104056 (2016) [arXiv:1609.01758].

\bibitem{Mazharimousavi:2021uki}
S.~H.~Mazharimousavi and M.~Halilsoy,
Annals Phys. \textbf{433}, 168579 (2021)
doi:10.1016/j.aop.2021.168579

\bibitem{Shabad2}A. E. Shabad, V. V. Usov, Phys. Rev. D \textbf{83}, 105006 (2011) [arXiv:1101.2343].

\bibitem{Managave:2023rhn}
K.~G.~Managave, H.~A.~Redekar, R.~B.~Kumbhar, S.~P.~Das and K.~Y.~Rajpure,
[arXiv:2303.07736 [gr-qc]].

\bibitem{Bronnikov} K. A. Bronnikov, Phys. Rev. D \textbf{63}, 044005 (2001).

\end{thebibliography}
\end{document}